\documentclass[aps,twocolumn,pra,showpacs,amsmath,amssymb,showkeys,10pt]{revtex4-2}
\usepackage{graphicx}
\usepackage{epstopdf}
\pdfoutput=1
\usepackage{braket}
\usepackage{hyperref}
\usepackage{longtable}

\begin{document}

	\title{Simulating and comparing the quantum and classical mechanical motion of two hydrogen atoms}
	
	\author{Hui-hui Miao}
	\email[Correspondence to: Vorobyovy Gory 1, Moscow, 119991, Russia. E-mail address: ]{hhmiao@cs.msu.ru (H.-H. Miao)}
	\affiliation{Faculty of Computational Mathematics and Cybernetics, Lomonosov Moscow State University, Vorobyovy Gory 1, Moscow, Russia}

	\date{\today}

	\begin{abstract}	
	This work presents a comprehensive comparison of quantum-mechanical and classical evolution for nuclear motion within a finite-dimensional quantum chemistry model. We employ a modified Tavis--Cummings--Hubbard model featuring two two-level artificial atoms in optical cavities to simulate the association and dissociation of a neutral hydrogen molecule. The initial conditions leading to the formation and decomposition of the molecule are examined. Dissipation in a Markovian open system is simulated by solving the Lindblad master equation. We compare the quantum and classical descriptions of nuclear motion: quantum mobility is characterized by nuclear tunneling, while classical motion is represented via fluctuations in interaction strengths. The emergence of dark states during dissociation and their significant impact on the evolutionary outcome are also analyzed. Key findings show that classical motion reaches the same final molecular states as quantum tunneling but requires an order-of-magnitude shorter time, with distinct patterns of population evolution. Two singlet dark states are identified that modify the branching ratio between neutral and ionic products.
	\end{abstract}


	\maketitle

\section{Introduction}
\label{sec:Intro}

The application of mathematical modeling in chemical prediction is continuously expanding, and its theoretical foundation has been elucidated in numerous studies \cite{Zhu2020, Ozhigov2021, Wang2021, McClean2021, Claudino2022, ChenYou2022, Miao2023, MiaoOzhigov2023, You2025}. In chemical modeling research, hydrogen-related processes have long been a focus of attention \cite{Zhu2020, Ozhigov2021, ChenYou2022, Miao2023, MiaoOzhigov2023}, especially the dynamic behavior of the neutral hydrogen molecule $\text{H}_2$ \cite{Miao2023, MiaoOzhigov2023}. This paper constructs a mathematical model describing the formation and dissociation of neutral hydrogen molecules. It focuses on the initial conditions that trigger molecule formation and decomposition, and compares the modes of motion of the two atomic nuclei under the quantum mechanical framework and the classical approximation.

Quantum electrodynamics (QED) has laid a unique theoretical foundation for exploring the interaction between light and matter, giving rise to a series of classical theoretical frameworks, including the quantum Rabi model (QRM) \cite{Rabi1936, Rabi1937}, the Dicke model \cite{Dicke1954}, the Hopfield model \cite{Hopfield1958}, the Jaynes--Cummings model (JCM) \cite{Jaynes1963}, the Tavis--Cummings model (TCM) \cite{Tavis1968}, and related models for ultra-strong coupling (USC) and deep strong coupling (DSC) mechanisms \cite{Casanova2010, Haroche2013, Gu2017, Forn-Diaz2019, Kockum2019, KockumMiranowicz2019}. The core theoretical basis of this paper is derived from the Tavis--Cummings--Hubbard model (TCHM) \cite{Angelakis2007}, which extends the TCM to multi-cavity systems interconnected by optical fibers. TCHM represents one of the simplest strong coupling (SC) models, operating in the regime
\begin{equation}
	\label{eq:Eta}
	\eta = \max\left( \frac{g}{\hbar\omega_c}, \frac{g}{\hbar\omega_a} \right) < 0.1,
\end{equation}
where $\hbar = h/2\pi$ is the reduced Planck constant (Dirac constant), $g$ is the coupling strength, $\omega_c$ is the cavity frequency, and $\omega_a$ is the atomic transition frequency. Due to its relative experimental accessibility, the SC regime has attracted considerable recent attention, with studies covering phase transitions \cite{Prasad2018, Wei2021}, quantum many-body phenomena \cite{Smith2021}, quantum gates \cite{OzhigovYI2020, Dull2021}, entropy \cite{Miao2024}, quantum discord \cite{MiaoLi2025}, dark states \cite{Lee1999, Andre2002, Poltl2012, Tanamoto2012, Hansom2014, Kozyrev2018, Ozhigov2019, Ozhigov2020}, among other topics \cite{Guo2019, Victorova2020, Kulagin2022, Afanasyev2022, Pluzhnikov2022, MiaoOzhigov2024, LiMiao2024}.

This paper aims to compare the evolutionary behavior of atomic nuclei under the quantum mechanical and classical mechanical descriptions. We obtain the Markov dissipative dynamics of the system by solving the quantum master equation (QME) \cite{Alicki1979, Breuer2002, Kosloff2013}, with dark-state characteristics being another core research topic. Specifically, the following research questions are addressed:
\begin{enumerate}
    \item Under the same initial conditions, what are the quantitative differences in the association/dissociation dynamics of $\text{H}_2$ between quantum nuclear tunneling and classical nuclear motion described by time-dependent coupling strengths?
    \item How do dark states emerge during dissociation, and what role do they play in shaping the final product distribution?
    \item How does Markovian dissipation (Lindblad master equation) affect the stability of the final molecular/atomic states under the two motion regimes?
\end{enumerate}
By answering these questions, this work advances the fundamental understanding of quantum-classical correspondence in molecular systems and demonstrates how a finite-dimensional QED model can capture essential features of hydrogen chemistry.

The paper is structured as follows: Sec. \ref{subsec:Target} first introduces the theoretical model; Sec. \ref{subsec:QME} gives the specific form of the quantum master equation; Sec. \ref{subsec:Thermally} discusses the coupling effect between the system and its external environment; and Sec. \ref{subsec:QuanClassMotion} focuses on the comparison between the two motion modes. Based on this, Sec. \ref{sec:Method} provides a numerical method for solving the master equation, and Sec. \ref{sec:Simulation} presents the corresponding numerical simulation results. Finally, Sec. \ref{sec:ConcluFuture} summarizes the entire paper and looks forward to future research directions. Appendices supplement the relevant technical details.

\section{Association--dissociation model of the neutral hydrogen molecule}
\label{sec:AssDissModel}
	
\subsection{Target model}
\label{subsec:Target}

\begin{figure*}[th]
	\centerline{\includegraphics[width=1.\textwidth]{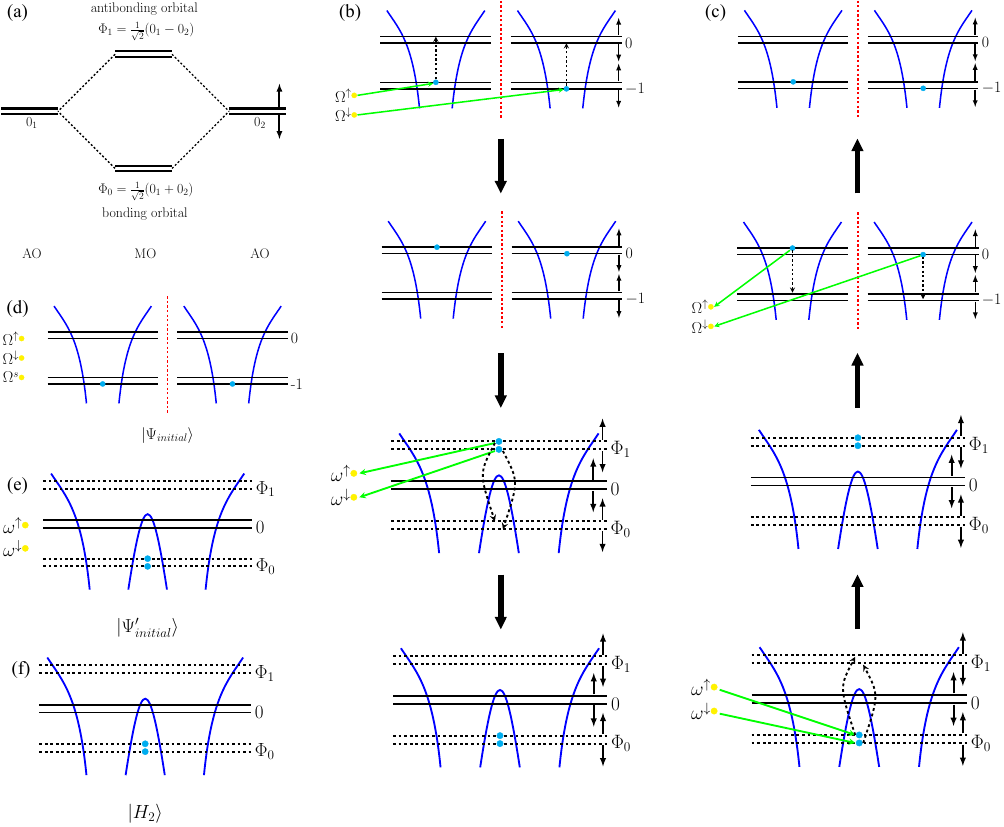}}
	\vspace*{8pt}
	\caption{(Color online) Association--dissociation model of the neutral hydrogen molecule. Color coding: blue dots = electrons; yellow dots = photons; red dashed lines = larger internuclear distances. Panel (a): atomic excited-state orbitals ($0_1$, $0_2$) hybridize to form the molecular bonding orbital $\Phi_0$ (ground) and antibonding orbital $\Phi_1$ (excited). Panel (b): association process. Two electrons with opposite spins in ground-state orbitals of different atoms absorb $\Omega^{\uparrow,\downarrow}$ photons (atomic transition modes) and transition to excited-state orbitals. Quantum tunneling brings the two nuclei into the same cavity, lowering the potential barrier. The two excited electrons enter $\Phi_1$ via hybridization, then rapidly emit $\omega^{\uparrow,\downarrow}$ photons (molecular transition modes) and relax to $\Phi_0$, forming a stable hydrogen molecule. Panel (c): dissociation process --- the reverse of panel (b). Panel (d): initial state for the association process, $|\Psi_{\text{initial}}\rangle$ --- two atoms in different cavities with $\Omega^{\uparrow,\downarrow,s}$ photons injected. Panel (e): initial state for the dissociation process, $|\Psi_{\text{initial}}'\rangle$ --- one hydrogen molecule in a single cavity with $\omega^{\uparrow,\downarrow}$ photons injected. Panel (f): final stable molecular state $|\mathrm{H}_2\rangle$. Symbols: $\Omega^{\uparrow,\downarrow,s}$ denote photon modes for atomic transitions (including spin flip), $\omega^{\uparrow,\downarrow}$ denote photon modes for molecular transitions.}
	\label{fig:AssDissModel}
\end{figure*}
	
Fig. \ref{fig:AssDissModel} illustrates the target model constructed in this paper. In this model, each energy level of atoms and molecules is further split into two sub-levels, corresponding to the spin-up ($\uparrow$) and spin-down ($\downarrow$) states, respectively. According to the Pauli exclusion principle \cite{Pauli1925}, a single sub-level can be occupied by at most one electron. Fig. \ref{fig:AssDissModel} (a) depicts the process by which atomic orbitals (AOs) hybridize to form molecular orbitals (MOs), where the molecular ground orbital (the bonding orbital) takes the form:
\begin{equation}
	\label{eq:MolStatePhi0}
	\Phi_0=\frac{1}{\sqrt{2}}\left(0_1+0_2\right).
\end{equation}
The molecular excited orbital (the antibonding orbital) takes the form:
\begin{equation}
	\label{eq:MolStatePhi1}
	\Phi_1=\frac{1}{\sqrt{2}}\left(0_1-0_2\right),
\end{equation}
where $\Phi_0$ is the ground-state molecular orbital, $\Phi_1$ is the excited-state molecular orbital, $0_1$ is the excited orbital of the first nucleus, and $0_2$ is the excited orbital of the second nucleus. Orbital hybridization can only occur in the excited state energy levels of atoms, while ground state orbitals (such as $-1_1$ and $-1_2$) do not meet this condition. In this model, electrons are bound in potential wells around the atomic nuclei. Fig. \ref{fig:AssDissModel} (b) depicts the process of two hydrogen atoms associating through orbital hybridization and covalent bond formation to form $\text{H}_2$; in contrast, Fig. \ref{fig:AssDissModel} (c) shows the reverse process of $\text{H}_2$ dissociating into hydrogen atoms, in which orbital dehybridization and covalent bond breaking occur simultaneously.

We employ the second quantization formalism \cite{Dirac1927, Fock1932} to construct the Hilbert space. As a result, the total Hilbert space takes the following form:
\begin{equation}
	\label{eq:SpaceC}
	\begin{aligned}
		&|\Psi\rangle_{\mathcal{C}}=|p_1\rangle_{\omega^{\uparrow}}|p_2\rangle_{\omega^{\downarrow}}|p_3\rangle_{\Omega^{\uparrow}}|p_4\rangle_{\Omega^{\downarrow}}|p_5\rangle_{\Omega^s}\\
		&\otimes|l_1\rangle_{\substack{\text{at}_1\\\text{or}_0}}^{\uparrow}|l_2\rangle_{\substack{\text{at}_1\\\text{or}_0}}^{\downarrow}|l_3\rangle_{\substack{\text{at}_1\\\text{or}_{-1}}}^{\uparrow}|l_4\rangle_{\substack{\text{at}_1\\\text{or}_{-1}}}^{\downarrow}|l_5\rangle_{\substack{\text{at}_2\\\text{or}_0}}^{\uparrow}|l_6\rangle_{\substack{\text{at}_2\\\text{or}_0}}^{\downarrow}|l_7\rangle_{\substack{\text{at}_2\\\text{or}_{-1}}}^{\uparrow}|l_8\rangle_{\substack{\text{at}_2\\\text{or}_{-1}}}^{\downarrow}|k\rangle_n,
	\end{aligned}
\end{equation}
where $p_{i\in\{1,2,\dots,5\}}$ denotes the number of photons in each mode: $\omega^{\uparrow}$ and $\omega^{\downarrow}$ are the photon modes corresponding to electron transitions between molecular orbitals, $\Omega^{\uparrow}$ and $\Omega^{\downarrow}$ are the photon modes for electron transitions between atomic orbitals, and $\Omega^{s}$ is the photon mode for spin-flips in the atomic system. In this paper, we assume that all photon modes have sufficiently large wavelengths to interact with electrons in any cavity. $l_{i\in\{1,2,\dots,8\}}$ denotes the orbital state: $l_i = 1$ indicates that the orbital is occupied by one electron, and $l_i = 0$ indicates that the orbital is unoccupied. The states of the nuclei are denoted by $|k\rangle_n$: $k = 0$ corresponds to both nuclei being localized in the same cavity, and $k = 1$ corresponds to the nuclei being distributed in different cavities.

The coupled-system Hamiltonian of the association--dissociation model with consideration of the rotating wave approximation (RWA; see Appx. \ref{appx:RWA}) is expressed by the total energy operator:
\begin{equation}
	\label{eq:Hamil}
	H=H_{\mathcal{A}}+H_{\mathcal{D}}+H_{\text{tun}}+H_{\text{spin}}.
\end{equation}
$H_{\mathcal{A}}$ denotes the associative Hamiltonian and takes the following form:
\begin{equation}
	\label{eq:HamilA}
	\begin{aligned}
		H_{\mathcal{A}}&=\left\{\hbar\omega^{\uparrow}a_{\omega^{\uparrow}}^{\dag}a_{\omega^{\uparrow}}+\hbar\omega^{\downarrow}a_{\omega^{\downarrow}}^{\dag}a_{\omega^{\downarrow}}\right.\\
		&+\hbar\omega^{\uparrow}\sigma_{\omega^{\uparrow}}^{\dag}\sigma_{\omega^{\uparrow}}+\hbar\omega^{\downarrow}\sigma_{\omega^{\downarrow}}^{\dag}\sigma_{\omega^{\downarrow}}\\
		&+g_{\omega^{\uparrow}}\left(a_{\omega^{\uparrow}}^{\dag}\sigma_{\omega^{\uparrow}}+a_{\omega^{\uparrow}}\sigma_{\omega^{\uparrow}}^{\dag}\right)\\
		&\left.+g_{\omega^{\downarrow}}\left(a_{\omega^{\downarrow}}^{\dag}\sigma_{\omega^{\downarrow}}+a_{\omega^{\downarrow}}\sigma_{\omega^{\downarrow}}^{\dag}\right)\right\}\sigma_n\sigma_n^{\dag},
	\end{aligned}
\end{equation}
where $\sigma_n\sigma_n^{\dag}$ indicates that the two nuclei are in close proximity. $g_{\omega^{\uparrow,\downarrow}}$ is the coupling strength between the photon mode $\omega^{\uparrow,\downarrow}$ (with annihilation and creation operators $a_{\omega^{\uparrow,\downarrow}}$ and $a_{\omega^{\uparrow,\downarrow}}^{\dag}$, respectively) and the electrons in the molecule (with excitation and relaxation operators $\sigma_{\omega^{\uparrow,\downarrow}}^{\dag}$ and $\sigma_{\omega^{\uparrow,\downarrow}}$, respectively). $H_{\mathcal{D}}$ denotes the dissociative Hamiltonian and takes the following form:
\begin{equation}
	\label{eq:HamilD}
	\begin{aligned}
		H_{\mathcal{D}}&=\left\{\hbar\Omega^{\uparrow}a_{\Omega^{\uparrow}}^{\dag}a_{\Omega^{\uparrow}}+\hbar\Omega^{\downarrow}a_{\Omega^{\downarrow}}^{\dag}a_{\Omega^{\downarrow}}\right.\\
		&+\sum_{i=1,2}\left(\hbar\Omega^{\uparrow}\sigma_{\Omega^{\uparrow},i}^{\dag}\sigma_{\Omega^{\uparrow},i}+\hbar\Omega^{\downarrow}\sigma_{\Omega^{\downarrow},i}^{\dag}\sigma_{\Omega^{\downarrow},i}\right)\\
		&+\sum_{i=1,2}\left[g_{\Omega^{\uparrow}}\left(a_{\Omega^{\uparrow}}^{\dag}\sigma_{\Omega^{\uparrow},i}+a_{\Omega^{\uparrow}}\sigma_{\Omega^{\uparrow},i}^{\dag}\right)\right.\\
		&\left.\left.+g_{\Omega^{\downarrow}}\left(a_{\Omega^{\downarrow}}^{\dag}\sigma_{\Omega^{\downarrow},i}+a_{\Omega^{\downarrow}}\sigma_{\Omega^{\downarrow},i}^{\dag}\right)\right]\right\}\sigma_n^{\dag}\sigma_n,
	\end{aligned}
\end{equation}
where the operator product $\sigma_n^{\dag}\sigma_n$ indicates that the two nuclei are far apart, and $i$ denotes the atomic index. Similarly, $g_{\Omega^{\uparrow,\downarrow}}$ is the coupling strength between the photon mode $\Omega^{\uparrow,\downarrow}$ and the electrons in the atom. $H_{\text{tun}}$ denotes the quantum tunneling effect between $H_{\mathcal{A}}$ and $H_{\mathcal{D}}$, and takes the following form:
\begin{equation}
	\label{eq:HamilTunnel}
	\begin{aligned}
		&H_{\text{tun}}=\left(\zeta_2\sigma_{\omega^{\uparrow}}^{\dag}\sigma_{\omega^{\uparrow}}\sigma_{\omega^{\downarrow}}^{\dag}\sigma_{\omega^{\downarrow}}+\zeta_1\sigma_{\omega^{\uparrow}}\sigma_{\omega^{\uparrow}}^{\dag}\sigma_{\omega^{\downarrow}}^{\dag}\sigma_{\omega^{\downarrow}}\right.\\
		&\left.+\zeta_1\sigma_{\omega^{\uparrow}}^{\dag}\sigma_{\omega^{\uparrow}}\sigma_{\omega^{\downarrow}}\sigma_{\omega^{\downarrow}}^{\dag}+\zeta_0\sigma_{\omega^{\uparrow}}\sigma_{\omega^{\uparrow}}^{\dag}\sigma_{\omega^{\downarrow}}\sigma_{\omega^{\downarrow}}^{\dag}\right)\left(\sigma_n^{\dag}+\sigma_n\right),
	\end{aligned}
\end{equation}
where $\sigma_{\omega^{\uparrow}}^{\dag}\sigma_{\omega^{\uparrow}}\sigma_{\omega^{\downarrow}}^{\dag}\sigma_{\omega^{\downarrow}}$ corresponds to both electrons (with opposite spins) occupying the orbital $\Phi_1$, with a large tunneling intensity $\zeta_2$; $\sigma_{\omega^{\uparrow}}\sigma_{\omega^{\uparrow}}^{\dag}\sigma_{\omega^{\downarrow}}^{\dag}\sigma_{\omega^{\downarrow}}$ corresponds to the spin-$\uparrow$ electron in $\Phi_0$ and the spin-$\downarrow$ electron in $\Phi_1$, with a low tunneling intensity $\zeta_1$; $\sigma_{\omega^{\uparrow}}^{\dag}\sigma_{\omega^{\uparrow}}\sigma_{\omega^{\downarrow}}\sigma_{\omega^{\downarrow}}^{\dag}$ corresponds to the spin-$\uparrow$ electron in $\Phi_1$ and the spin-$\downarrow$ electron in $\Phi_0$, also with a low tunneling intensity $\zeta_1$; $\sigma_{\omega^{\uparrow}}\sigma_{\omega^{\uparrow}}^{\dag}\sigma_{\omega^{\downarrow}}\sigma_{\omega^{\downarrow}}^{\dag}$ corresponds to both electrons occupying the orbital $\Phi_0$, with a tunneling intensity $\zeta_0 = 0$. The quantum tunneling effect is reduced when an electron transitions to the molecular ground state. $H_{\text{spin}}$ denotes the electron spin transition and takes the following form:
\begin{equation}
	\label{eq:HamilSpin}
	\begin{aligned}
		H_{\text{spin}}&=\hbar\Omega^sa_{\Omega^s}^{\dag}a_{\Omega^s}+\hbar\Omega^s\sum_{i=1,2}\sigma_{\Omega^s,i}^{\dag}\sigma_{\Omega^s,i}\\
		&+g_{\Omega^s}\sum_{i=1,2}\left(a_{\Omega^s}^{\dag}\sigma_{\Omega^s,i}+a_{\Omega^s}\sigma_{\Omega^s,i}^{\dag}\right),
	\end{aligned}
\end{equation}
where $g_{\Omega^s}$ is the coupling strength between the photon mode $\Omega^s$ and the electrons in the atom. This paper introduces a spin-photon model containing the mode $\Omega^s$ to describe the association--dissociation process of the atomic nuclei. This model allows transitions between spins $\uparrow$ and $\downarrow$, but such transitions are strictly constrained by the Pauli exclusion principle \cite{Pauli1925}: two electrons with the same spin are not allowed at the same energy level. We further point out that the occurrence of electron spin transitions requires specific conditions --- the process is only allowed when the electron is in the atomic state $|1\rangle_n$; conversely, if the electron is in the molecular state $|0\rangle_n$, spin transitions are prohibited because they violate the Pauli principle.

For the definitions and derivations of all operators involved in Eqs. \eqref{eq:HamilA}--\eqref{eq:HamilSpin}, please refer to Appx. \ref{appx:Operators}.
	
\subsection{Quantum master equation} 
\label{subsec:QME}
	
The dynamics of the system are described by solving the QME in the Markovian approximation for the density operator $\rho$:
\begin{equation}
	\label{eq:QME}
	i\hbar\dot{\rho}=\left[H,\rho\right]+iL\left(\rho\right),
\end{equation}
where $\left[H,\rho\right]=H\rho-\rho H$. We consider a graph $\mathcal{K}$ of the potential photon dissipation channels between states. The edges and vertices of $\mathcal{K}$ represent the permitted dissipations and the states, respectively. Similarly, $\mathcal{K}'$ is a graph of potential photon influxes. $L\left(\rho\right)$ is as follows:
\begin{equation}
	\label{eq:LindbladOperator}
	L\left(\rho\right)=\sum_{k\in \mathcal{K}} L_k\left(\rho\right)+\sum_{k'\in \mathcal{K}'} L_{k'}\left(\rho\right),
\end{equation}
where $L_k\left(\rho\right)$ is the standard dissipation superoperator corresponding to the jump operator $A_k$ and taking the density matrix $\rho$ as its argument:
\begin{equation}
	\label{eq:DissSuper}
	L_k\left(\rho\right)=\gamma_k\left(A_k\rho A_k^{\dag}-\frac{1}{2}\left\{\rho, A_k^{\dag}A_k\right\}\right),
\end{equation}
where $\left\{\rho, A_k^{\dag}A_k\right\}=\rho A_k^{\dag}A_k + A_k^{\dag}A_k\rho$. The term $\gamma_k$ refers to the overall spontaneous emission rate (dissipation intensity) for photons in the channels $k\in \mathcal{K}$, arising from photon leakage from the cavity to the external environment. Similarly, $L_{k'}\left(\rho\right)$ is the standard influx superoperator, which takes the following form:
\begin{equation}
	\label{eq:InfluxSuper}
	L_{k'}\left(\rho\right)=\gamma_{k'}\left(A_k^{\dag}\rho A_k-\frac{1}{2}\left\{\rho, A_kA_k^{\dag}\right\}\right),
\end{equation}
where the total spontaneous influx rate (influx intensity) for photons in the channels $k'\in \mathcal{K}'$ is denoted by $\gamma_{k'}$.

\subsection{Thermally stationary state} 
\label{subsec:Thermally}
		
In this subsection, we introduce the relationship between the dissipation intensity and the influx intensity. The stationary state of a field at temperature $\mathcal{T}$ is defined as a mixed state with a Gibbs distribution of Fock components:
\begin{equation}
	\label{eq:Photon_gibbs}
	{\cal G}\left(\mathcal{T}\right)_f=c \sum\limits_{p=0}^\infty \exp\left(-\frac{\hbar\omega p}{K\mathcal{T}}\right)|p\rangle\langle p|,
\end{equation} 
where $K$ is the Boltzmann constant, $c$ is the normalization factor, $p$ is the number of photons, and $\omega$ is the photonic mode. The notation $\gamma_{k'}/\gamma_{k}=\mu$ is introduced. Since the temperature would otherwise be infinitely high and the state ${\cal G}\left(\mathcal{T}\right)_f$ would not be normalizable, the state exists only at $\mu<1$ \cite{Kulagin2019}. The probability of the photonic Fock state $|p\rangle$ at temperature $\mathcal{T}$ is proportional to $\exp\left(-\frac{\hbar\omega}{K\mathcal{T}}\right)$. We assume that
\begin{equation}
	\label{eq:PopulationFock}
	\mu=\exp\left(-\frac{\hbar\omega}{K\mathcal{T}}\right),
\end{equation}
which leads to $\mathcal{T}=\frac{\hbar\omega}{K\ln\left(1/\mu\right)}$.
	
\subsection{Quantum and classical motion of nuclei} 
\label{subsec:QuanClassMotion}
	
\begin{figure*}[th]
	\centerline{\includegraphics[width=1.\textwidth]{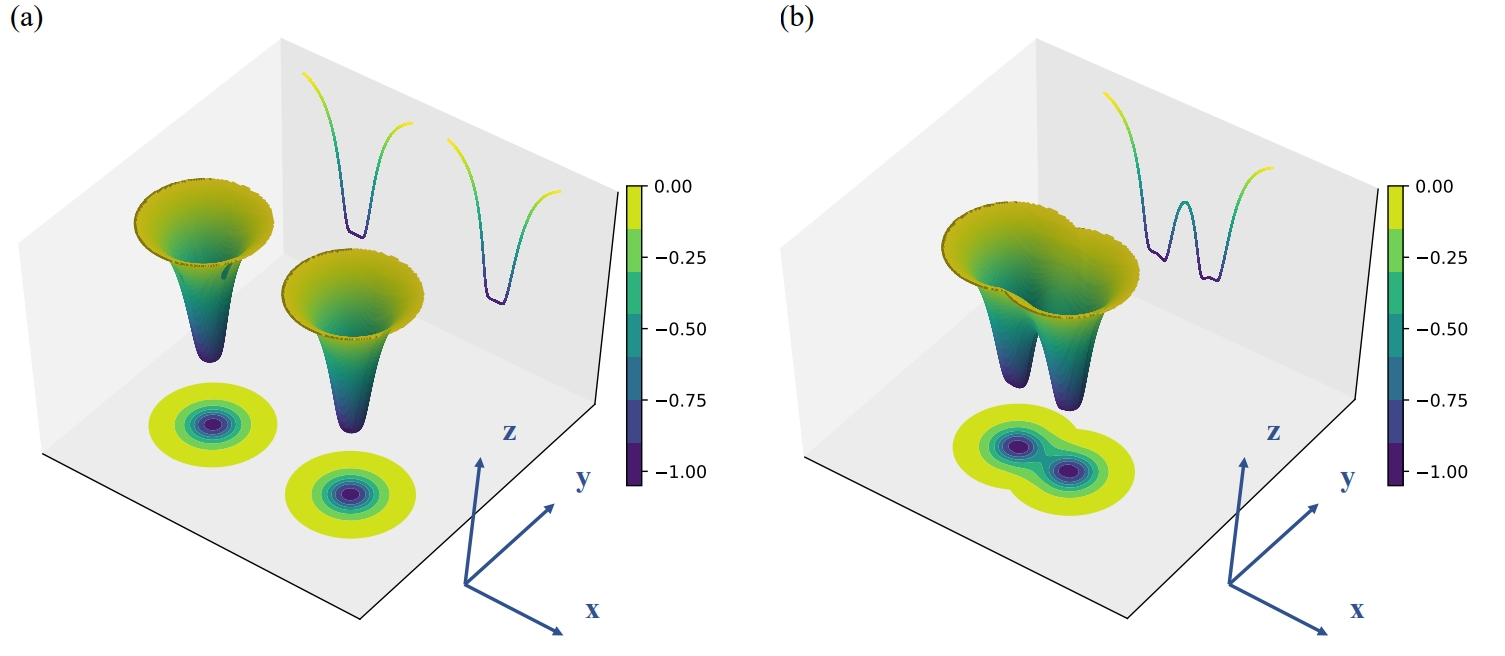}}
	\vspace*{8pt}
	\caption{(Color online) 3D diagram of the double potential well. Panel (a): double potential well for dissociative system --- nuclei far apart (different cavities). Panel (b): single potential well for associative system --- nuclei close together (same cavity). Two types of nuclear motion are considered in this work: (i) Quantum motion: instantaneous tunneling between (a) and (b); (ii) Classical motion: slow movement along the reaction coordinate from (a) to (b) (association process) or from (b) to (a) (dissociation process).}
	\label{fig:PotentialWell}
\end{figure*}
	
The three-dimensional surface diagram, shown in Fig. \ref{fig:PotentialWell}, intuitively shows how the double potential well around nuclei alters depending on how close or how far away they are. We propose two ways in which hydrogen nuclei can move:
\begin{itemize}
	\item Quantum motion of nuclei. In this regime, the nuclei utilize the quantum tunneling effect to move instantaneously between the optical cavities. As shown in Fig. \ref{fig:PotentialWell} (a), the two nuclei are located in different optical cavities. At this point, the orbitals associated with each nucleus remain atomic, and all interaction forces involving molecular orbitals are absent, i.e., $g_{\omega^{\uparrow,\downarrow}}=0$, while only the interaction forces involving atomic orbitals are present, i.e., $g_{\Omega^{\uparrow,\downarrow,s}}=g_{\Omega^{\uparrow,\downarrow,s}}^{\max}$. Here, $g_{\Omega^{\uparrow,\downarrow,s}}^{\max}$ is the maximum value of the interaction forces $g_{\Omega^{\uparrow,\downarrow,s}}(t)$ when the two nuclei are far enough apart that the system consists of two completely independent atoms. As shown in Fig. \ref{fig:PotentialWell} (b), the two nuclei are located in the same cavity. At this point, the orbitals remain molecular, and all interaction forces involving atomic orbitals are absent, i.e., $g_{\Omega^{\uparrow,\downarrow,s}}=0$, while only the interaction forces involving molecular orbitals are present, i.e., $g_{\omega^{\uparrow,\downarrow}}=g_{\omega^{\uparrow,\downarrow}}^{\max}$. Here, $g_{\omega^{\uparrow,\downarrow}}^{\max}$ is the maximum value of the interaction forces $g_{\omega^{\uparrow,\downarrow}}(t)$ when the two nuclei are close enough that the system consists of a single molecule. Due to the instantaneous nature of the quantum tunneling effect (from panel (a) to panel (b) or vice versa), the change in the strength of the interaction force occurs instantaneously. Quantum tunneling enables instantaneous transitions between the two potential well configurations, corresponding to an abrupt change in coupling strengths.
	\item Classical motion of nuclei. In this regime, the movement of the nuclei is not instantaneous but occurs slowly. When the nuclei slowly approach each other, the whole system gradually evolves from two independent atoms into a molecule. This process is accompanied by the interaction forces $g_{\Omega^{\uparrow,\downarrow,s}}$ gradually decreasing from $g_{\Omega^{\uparrow,\downarrow,s}}^{\max}$ to $0$, while the interaction forces $g_{\omega^{\uparrow,\downarrow}}$ gradually increase from $0$ to $g_{\omega^{\uparrow,\downarrow}}^{\max}$. Conversely, when the nuclei slowly move apart, the whole system gradually evolves from a molecule into two independent atoms. This process is accompanied by the interaction forces $g_{\omega^{\uparrow,\downarrow}}$ gradually decreasing from $g_{\omega^{\uparrow,\downarrow}}^{\max}$ to $0$, while the interaction forces $g_{\Omega^{\uparrow,\downarrow,s}}$ gradually increase from $0$ to $g_{\Omega^{\uparrow,\downarrow,s}}^{\max}$. Obviously, the magnitude of the interaction force is related to the distance between the nuclei. We consider two scenarios:
\begin{figure*}[th]
	\centerline{\includegraphics[width=.85\textwidth]{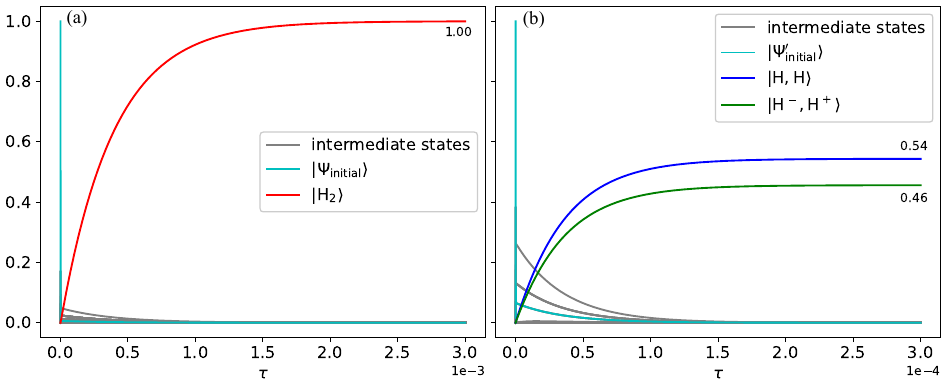}}
	\vspace*{8pt}
	\caption{(Color online) Quantum motion of the nuclei (tunneling regime). All curves show state populations as functions of time $\tau$ (horizontal axis). Panel (a): association process starting from the initial state $|\Psi_{\text{initial}}\rangle$ (Fig. \ref{fig:AssDissModel} (d)). Red curve: population of the stable molecular state $|\mathrm{H}_2\rangle$ (Fig. \ref{fig:AssDissModel} (f)). The population reaches unity at $\tau=T$ ($T$ is the total time), indicating complete conversion of two free atoms into one hydrogen molecule. Panel (b): dissociation process starting from the initial state $|\Psi_{\text{initial}}'\rangle$ (Fig. \ref{fig:AssDissModel} (e)). Blue curve: neutral atomic pair $|\mathrm{H},\mathrm{H}\rangle$ (see Fig. \ref{fig:FinalStates} for state components). Green curve: ionic pair $|\mathrm{H}^-,\mathrm{H}^+\rangle$ (also in Fig. \ref{fig:FinalStates}). Asymptotic populations at $\tau=T$ are $\approx 0.544$ (blue) and $\approx 0.456$ (green). Parameters: $\mu_{\Omega^{\uparrow,\downarrow,s}}=0.5,\mu_{\omega^{\uparrow,\downarrow}}=0$ for panel (a) and $\mu_{\Omega^{\uparrow,\downarrow,s}}=0,\mu_{\omega^{\uparrow,\downarrow}}=0.5$ for panel (b).}
	\label{fig:QuantumMotion}
\end{figure*}
\begin{figure}[th]
	\centerline{\includegraphics[width=0.5\textwidth]{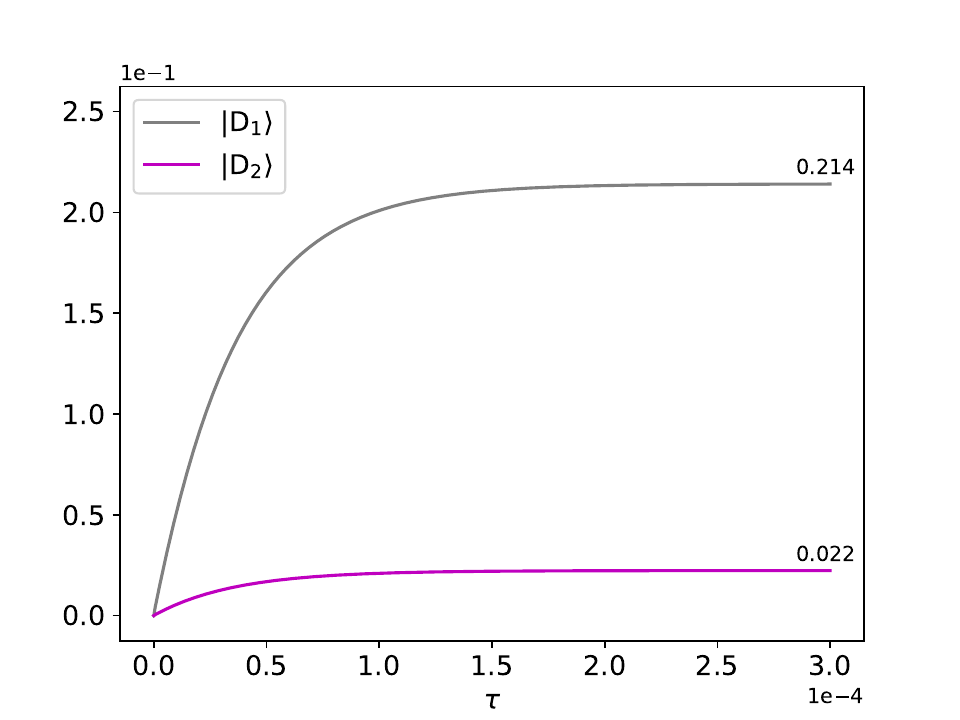}}
	\vspace*{8pt}
	\caption{(Color online) Dark states in the quantum motion regime during dissociation. Two singlet dark state combinations are identified: $|\text{D}_1\rangle = \frac{1}{\sqrt{2}}(|\Psi_0\rangle-|\Psi_1\rangle)$ and $|\text{D}_2\rangle = \frac{1}{\sqrt{2}}(|\Psi_2\rangle-|\Psi_3\rangle)$, which satisfy $a_{\Omega^s}^{\dag}\bar{\sigma}_{\Omega^s}|\text{D}_1\rangle=0$ and $a_{\Omega^{\downarrow}}^{\dag}\bar{\sigma}_{\Omega^{\downarrow}}|\text{D}_2\rangle=0$ (see Fig. \ref{fig:Combinations} for the formation mechanism). Their populations during dissociation are $0.214$ and $0.022$, respectively. The same dark states also appear in the classical motion case (Fig. \ref{fig:DarkStatesClassMotion}).}
	\label{fig:DarkStatesQuanMotion}
\end{figure}
	\begin{itemize}
		\item We first consider a simple scenario in which the two nuclei approach each other at a constant velocity. In this case, the value of $g_{\Omega^{\uparrow,\downarrow,s}}$ decreases uniformly over time $\tau$, while the value of $g_{\omega^{\uparrow,\downarrow}}$ increases uniformly over time $\tau$. The $\tau$-dependent curves of $g_{\Omega^{\uparrow,\downarrow,s}}$ and $g_{\omega^{\uparrow,\downarrow}}$ during the association process are shown below:
		\begin{subequations}
			\label{eq:StraightAss}
			\begin{align}
				g_{\Omega^{\uparrow,\downarrow,s}}(\tau)&=g_{\Omega^{\uparrow,\downarrow,s}}^{\max}\left(1-\frac{\tau}{T}\right),\label{eq:StraightAssAts}\\
				g_{\omega^{\uparrow,\downarrow}}(\tau)&=g_{\omega^{\uparrow,\downarrow}}^{\max}\frac{\tau}{T},\label{eq:StraightAssMol}
			\end{align}
		\end{subequations}
		where $T$ is the total time required to complete the evolution. Similarly, we define the corresponding equations for the dissociation process (the reverse of the association process) as follows:
		\begin{subequations}
			\label{eq:StraightDiss}
			\begin{align}
				g_{\Omega^{\uparrow,\downarrow,s}}(\tau)&=g_{\Omega^{\uparrow,\downarrow,s}}^{\max}\frac{\tau}{T},\label{eq:StraightDissAts}\\
				g_{\omega^{\uparrow,\downarrow}}(\tau)&=g_{\omega^{\uparrow,\downarrow}}^{\max}\left(1-\frac{\tau}{T}\right).\label{eq:StraightDissMol}
			\end{align}
		\end{subequations}
		\item Now we consider a more complex yet more realistic scenario in which the classical motion of the nuclei is non-uniform. Assume that the two nuclei approach each other, with their speed initially increasing until a certain critical point (here taken as the midpoint) and then decreasing to zero due to mutual repulsion after passing that point. In this case, the rate at which $g_{\Omega^{\uparrow,\downarrow,s}}$ decays first increases and then decreases until $g_{\Omega^{\uparrow,\downarrow,s}}$ reaches zero, while the rate at which $g_{\omega^{\uparrow,\downarrow}}$ grows also first increases and then decreases until $g_{\omega^{\uparrow,\downarrow}}$ reaches its maximum value. We assume that the $\tau$-dependent curves of 
$g_{\Omega^{\uparrow,\downarrow,s}}$ and $g_{\omega^{\uparrow,\downarrow}}$ follow a trigonometric functional form:
		\begin{subequations}
			\label{eq:TrigonometricAss}
			\begin{align}
				g_{\Omega^{\uparrow,\downarrow,s}}(\tau)&=g_{\Omega^{\uparrow,\downarrow,s}}^{\max}\left(\frac{\cos\left(\frac{\tau\pi}{T}\right)+1}{2}\right),\label{eq:TrigonometricAssAts}\\
				g_{\omega^{\uparrow,\downarrow}}(\tau)&=g_{\omega^{\uparrow,\downarrow}}^{\max}\left(\frac{\sin\left(\frac{\tau\pi}{T}-\frac{\pi}{2}\right)+1}{2}\right).\label{eq:TrigonometricAssMol}
			\end{align}
		\end{subequations}
		Similarly, we define the corresponding equations for the dissociation process as follows:
		\begin{subequations}
			\label{eq:TrigonometricDiss}
			\begin{align}
				g_{\Omega^{\uparrow,\downarrow,s}}(\tau)&=g_{\Omega^{\uparrow,\downarrow,s}}^{\max}\left(\frac{\sin\left(\frac{\tau\pi}{T}-\frac{\pi}{2}\right)+1}{2}\right),\label{eq:TrigonometricDissAts}\\
				g_{\omega^{\uparrow,\downarrow}}(\tau)&=g_{\omega^{\uparrow,\downarrow}}^{\max}\left(\frac{\cos\left(\frac{\tau\pi}{T}\right)+1}{2}\right).\label{eq:TrigonometricDissMol}
			\end{align}
		\end{subequations}
	\end{itemize}
\end{itemize}

\section{Numerical method} 
\label{sec:Method}

\begin{figure*}[th]
	\centerline{\includegraphics[width=1.\textwidth]{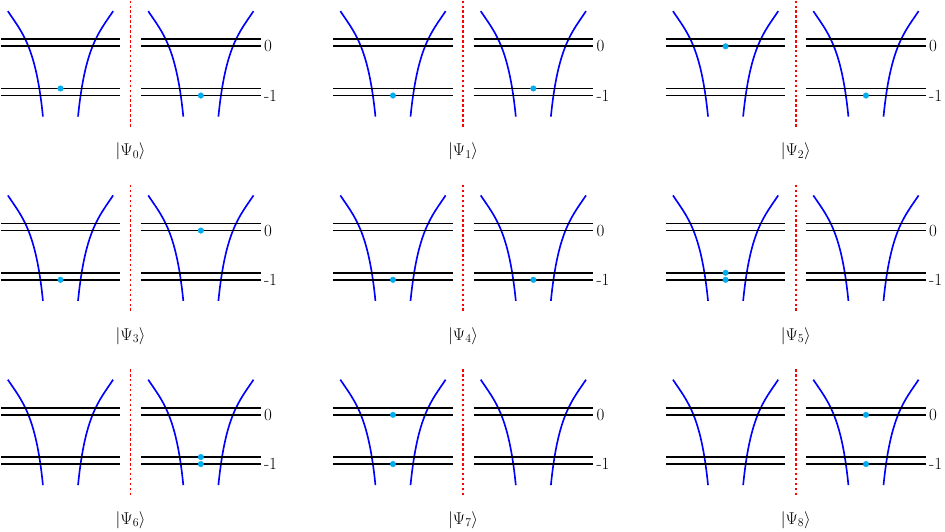}}
	\vspace*{8pt}
	\caption{(Color online) Nine final states in the dissociation process. The nine states $|\Psi_0\rangle$ through $|\Psi_8\rangle$ are divided into two types by net charge. $|\Psi_0\rangle$–$|\Psi_4\rangle$: each hydrogen atom carries one electron $\rightarrow$ neutral pair $|\mathrm{H},\mathrm{H}\rangle$ (blue curve in Fig. \ref{fig:QuantumMotion} (b), asymptotic population $\approx 0.544$). $|\Psi_5\rangle$–$|\Psi_8\rangle$: both electrons on the same hydrogen atom $\rightarrow$ ionic pair $|\mathrm{H}^-,\mathrm{H}^+\rangle$ (green curve in Fig. \ref{fig:QuantumMotion} (b), asymptotic population $\approx 0.456$).}
	\label{fig:FinalStates}
\end{figure*}

The solution $\rho(t)$ of Eq. \eqref{eq:QME} can be approximated by a sequence of two steps: in the first step, we advance by one step in solving the unitary part of Eq. \eqref{eq:QME}:
\begin{equation}
	\label{eq:UnitaryPart}
	\tilde{\rho}\left(t+dt\right)=e^{-\frac{i}{\hbar}\int_t^{t+dt}Hd\tau}\rho\left(t\right)e^{\frac{i}{\hbar}\int_t^{t+dt}Hd\tau},
\end{equation}
where $dt$ is the time step. We use integrals in Eq. \eqref{eq:UnitaryPart} because the Hamiltonian is $\tau$-dependent and the regions of equal energy within the Hamiltonian commute. We introduce the precise time-step integration method (PTSIM) to compute the matrix exponential (see Appx. \ref{appx:PTSIM}). In the second step, we advance by one step in solving Eq. \eqref{eq:QME} with the commutator removed:
\begin{equation}
	\label{eq:Solution}
	\rho\left(t+dt\right)=\tilde{\rho}\left(t+dt\right)+\frac{1}{\hbar}L\left(\tilde{\rho}(t+dt)\right)dt.
\end{equation}

\section{Simulations and results} 
\label{sec:Simulation}
	
Different from previous work, in this paper we study both the association and dissociation processes of the neutral hydrogen molecule. Thus, we consider two initial states:
\begin{itemize}
	\item The initial state $|\Psi_{\text{initial}}\rangle$, shown in Fig. \ref{fig:AssDissModel} (d), represents that the two nuclei are in different cavities, and the electrons are in the atomic ground state with spin down. We pump in two photons with different modes $\Omega^{\uparrow}$ and $\Omega^{\downarrow}$, along with a spin photon.
	\item The initial state $|\Psi_{\text{initial}}'\rangle$, shown in Fig. \ref{fig:AssDissModel} (e), represents that the two nuclei are in the same cavity, and the electrons are in the molecular ground state with different spins. We pump in only two photons with different modes $\omega^{\uparrow}$ and $\omega^{\downarrow}$.
\end{itemize}
	
\subsection{Results of quantum motion of nuclei} 
\label{subsec:ResultsQuantMotion}

\begin{figure*}[th]
	\centerline{\includegraphics[width=1.\textwidth]{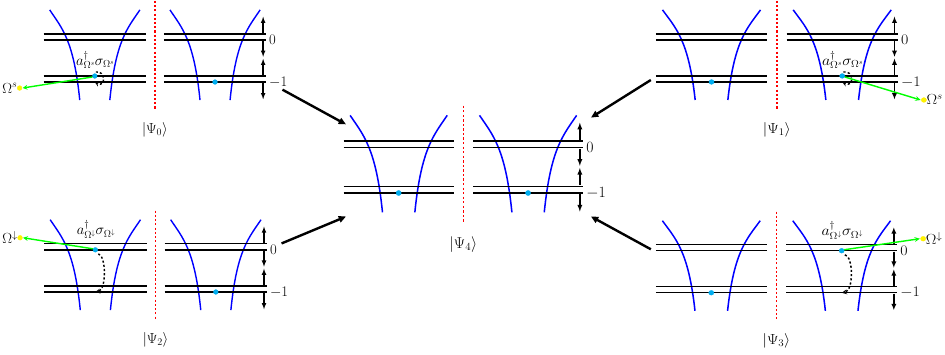}}
	\vspace*{8pt}
	\caption{(Color online) Formation mechanism of the two dark state singlets. Upper row: For $|\Psi_0\rangle$ and $|\Psi_1\rangle$, applying $a_{\Omega^s}^{\dag}\bar{\sigma}_{\Omega^s}$ flips the up-spin electron in the ground state to down-spin, yielding $|\Psi_4\rangle$ in both cases. The subtraction $\frac{1}{\sqrt{2}}(|\Psi_0\rangle-|\Psi_1\rangle)$ therefore gives $a_{\Omega^s}^{\dag}\bar{\sigma}_{\Omega^s}|\text{D}_1\rangle=0$, making $|\text{D}_1\rangle$ a dark state with respect to this dissipation channel. Lower row: For $|\Psi_2\rangle$ and $|\Psi_3\rangle$, applying $a_{\Omega^{\downarrow}}^{\dag}\bar{\sigma}_{\Omega^{\downarrow}}$ relaxes the excited-state electron to the ground state, also yielding $|\Psi_4\rangle$ in both cases. The subtraction $\frac{1}{\sqrt{2}}(|\Psi_2\rangle-|\Psi_3\rangle)$ gives $a_{\Omega^{\downarrow}}^{\dag}\bar{\sigma}_{\Omega^{\downarrow}}|\text{D}_2\rangle=0$, making $|\text{D}_2\rangle$ a dark state.}
	\label{fig:Combinations}
\end{figure*}
	
In simulations: $\hbar=1$, $\Omega^{\uparrow}=\Omega^{\downarrow}=10^{10}$, $\omega^{\uparrow}=\omega^{\downarrow}=5\times 10^9$, $\Omega^s=10^9$, $g_{\Omega^{\uparrow}}=g_{\Omega^{\downarrow}}=0.01\times\Omega^{\uparrow}$, $g_{\omega^{\uparrow}}=g_{\omega^{\downarrow}}=0.01\times\omega^{\uparrow}$, $g_{\Omega^s}=0.01\times\Omega^s$, $\zeta_2=10\times g_{\Omega^{\uparrow}}$, $\zeta_1=g_{\Omega^{\uparrow}}$, $\zeta_0=0$, and $\gamma_{\omega^{\uparrow}}=\gamma_{\omega^{\downarrow}}=\gamma_{\Omega^{\uparrow}}=\gamma_{\Omega^{\downarrow}}=\gamma_{\Omega^s}=0.1\times g_{\Omega^{\uparrow}}$.
	
Fig. \ref{fig:QuantumMotion} presents the results for quantum nuclear motion. With the initial state $|\Psi_{\text{initial}}\rangle$ in panel (a), only photons in the $\Omega^{\uparrow,\downarrow,s}$ modes are injected, while $\omega^{\uparrow,\downarrow}$ photon influx is suppressed, thereby promoting $\text{H}_2$ formation. As discussed in Sec. \ref{subsec:Thermally}, the influx rates are maintained below their corresponding dissipation rates, with the parameters set to $\mu_{\Omega^{\uparrow,\downarrow,s}}=0.5$ and $\mu_{\omega^{\uparrow,\downarrow}}=0$. The numerical simulation in Fig. \ref{fig:QuantumMotion} (a) shows that the population of the molecular state $|\mathrm{H}_2\rangle$ (red solid curve, defined in Fig. \ref{fig:AssDissModel} (f)) gradually increases, eventually reaching unity. This indicates complete conversion of free atoms into a stable $\text{H}_2$ molecule. For dissociation to occur, electrons would need to be excited from the molecular ground orbital via $\omega^{\uparrow,\downarrow}$ photons; however, the absence of these photon sources leads to their depletion through cavity leakage. Consequently, the system evolves toward a stable molecular configuration.
	
Panel (b) of Fig. \ref{fig:QuantumMotion} corresponds to the initial molecular state $|\Psi_{\text{initial}}'\rangle$. Here, only $\omega^{\uparrow,\downarrow}$ photon injection is permitted, while $\Omega^{\uparrow,\downarrow,s}$ modes are blocked, favoring $\text{H}_2$ dissociation. The parameters are set to $\mu_{\Omega^{\uparrow,\downarrow,s}}=0$ and $\mu_{\omega^{\uparrow,\downarrow}}=0.5$. The simulation results reveal two emerging final states: the neutral atomic pair $|\mathrm{H},\mathrm{H}\rangle$ (blue curve) and the ionic pair $|\mathrm{H}^-,\mathrm{H}^+\rangle$ (green curve), with asymptotic populations of $0.544$ and $0.456$, respectively. As detailed in Fig. \ref{fig:FinalStates}, $|\mathrm{H},\mathrm{H}\rangle$ comprises the components $|\Psi_0\rangle$ through $|\Psi_4\rangle$, while $|\mathrm{H}^-,\mathrm{H}^+\rangle$ consists of the components $|\Psi_5\rangle$ to $|\Psi_8\rangle$. The association process requires $\Omega^{\uparrow,\downarrow,s}$ photons to excite electrons from atomic ground orbitals; however, the absence of these photon sources results in their depletion via cavity leakage. The system thus evolves toward a final configuration of two independent hydrogen atoms, confirming successful $\text{H}_2$ dissociation.
	
During the dissociation of the hydrogen molecule, two dark state combinations (singlets, see Fig. \ref{fig:DarkStatesQuanMotion}) appear as follows:
\begin{subequations}
	\label{eq:DarkStates}
	\begin{align}
		|\text{D}_1\rangle&=\frac{1}{\sqrt{2}}\left(|\Psi_0\rangle-|\Psi_1\rangle\right),\label{eq:DarkState1}\\
		|\text{D}_2\rangle&=\frac{1}{\sqrt{2}}\left(|\Psi_2\rangle-|\Psi_3\rangle\right),\label{eq:DarkState2}
	\end{align}
\end{subequations}
The dark states $|\text{D}_1\rangle$ and $|\text{D}_2\rangle$ are decoupled from specific dissipation channels, which modifies the branching ratio between $|\mathrm{H},\mathrm{H}\rangle$ and $|\mathrm{H}^-,\mathrm{H}^+\rangle$. This also explains why the population of $|\mathrm{H},\mathrm{H}\rangle$ is greater than that of $|\mathrm{H}^-,\mathrm{H}^+\rangle$ in Fig. \ref{fig:QuantumMotion} (b). The formation mechanism of these two dark states is shown in Fig. \ref{fig:Combinations}, and we have:
\begin{subequations}
	\label{eq:Singlets}
	\begin{align}
		a_{\Omega^s}^{\dag}\bar{\sigma}_{\Omega^s}|\text{D}_1\rangle&=\frac{1}{\sqrt{2}}a_{\Omega^s}^{\dag}\bar{\sigma}_{\Omega^s}\left(|\Psi_0\rangle-|\Psi_1\rangle\right)\nonumber\\
		&=\frac{1}{\sqrt{2}}\left(a_{\Omega^s}^{\dag}\sigma_{\Omega^s,1}|\Psi_0\rangle-a_{\Omega^s}^{\dag}\sigma_{\Omega^s,2}|\Psi_1\rangle\right)\nonumber\\
		&=\frac{1}{\sqrt{2}}\left(|\Psi_4\rangle-|\Psi_4\rangle\right)\nonumber\\
		&=0,\label{eq:Singlet1}\\
		a_{\Omega^{\downarrow}}^{\dag}\bar{\sigma}_{\Omega^{\downarrow}}|\text{D}_2\rangle&=\frac{1}{\sqrt{2}}a_{\Omega^{\downarrow}}^{\dag}\bar{\sigma}_{\Omega^{\downarrow}}\left(|\Psi_2\rangle-|\Psi_3\rangle\right)\nonumber\\
		&=\frac{1}{\sqrt{2}}\left(a_{\Omega^{\downarrow}}^{\dag}\sigma_{\Omega^{\downarrow},1}|\Psi_2\rangle-a_{\Omega^{\downarrow}}^{\dag}\sigma_{\Omega^{\downarrow},2}|\Psi_3\rangle\right)\nonumber\\
		&=\frac{1}{\sqrt{2}}\left(|\Psi_4\rangle-|\Psi_4\rangle\right)\nonumber\\
		&=0,\label{eq:Singlet2}
	\end{align}
\end{subequations}
where $\bar{\sigma}_{\Omega^s}=\sigma_{\Omega^s,1}+\sigma_{\Omega^s,2}$ and $\bar{\sigma}_{\Omega^{\downarrow}}=\sigma_{\Omega^{\downarrow},1}+\sigma_{\Omega^{\downarrow},2}$.
	
\subsection{Results of classical motion of nuclei} 
\label{subsec:ResultsClassMotion}

\begin{figure*}[th]
	\centerline{\includegraphics[width=.85\textwidth]{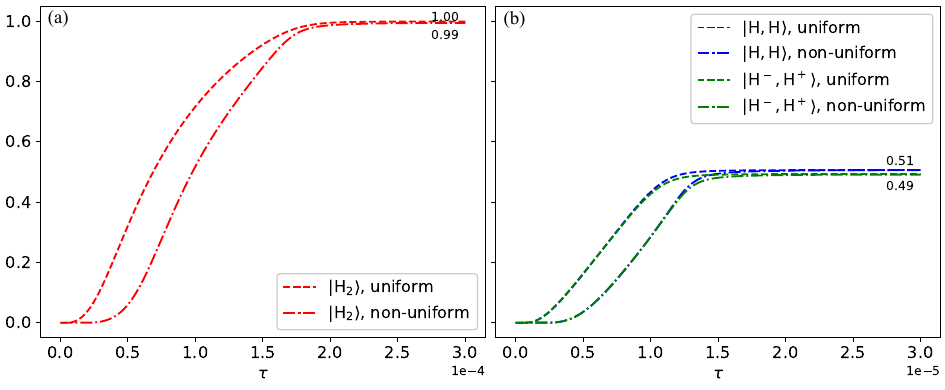}}
	\vspace*{8pt}
	\caption{(Color online) Classical motion of the nuclei. All curves show state populations as functions of time $\tau$ (horizontal axis). Panel (a): association process starting from $|\Psi_{\text{initial}}\rangle$ (Fig. \ref{fig:AssDissModel} (d)). Red curves: population of $|\mathrm{H}_2\rangle$ (Fig. \ref{fig:AssDissModel} (f)). Dashed red curve: uniform nuclear motion (Eq. \eqref{eq:StraightAss}). Dash-dotted red curve: non-uniform nuclear motion (Eq. \eqref{eq:TrigonometricAss}). Both reach unity at $\tau=T$, but uniform motion yields a faster initial rise. Panel (b): dissociation process starting from $|\Psi_{\text{initial}}'\rangle$ (Fig. \ref{fig:AssDissModel} (e)). Blue curves: neutral pair $|\mathrm{H},\mathrm{H}\rangle$; green curves: ionic pair $|\mathrm{H}^-,\mathrm{H}^+\rangle$ (final states detailed in Fig. \ref{fig:FinalStates}). Dashed curves: uniform motion (Eq. \eqref{eq:StraightDiss}). Dash-dotted curves: non-uniform motion (Eq. \eqref{eq:TrigonometricDiss}). Asymptotic populations at $\tau=T$ are $\approx 0.51$ (blue) and $\approx 0.49$ (green). Uniform motion also yields a faster initial rise in the dissociation process. Parameters are the same as in Fig. \ref{fig:QuantumMotion}, with $\mu_{\Omega^{\uparrow,\downarrow,s}}=0.5,\mu_{\omega^{\uparrow,\downarrow}}=0$ for panel (a) and $\mu_{\Omega^{\uparrow,\downarrow,s}}=0,\mu_{\omega^{\uparrow,\downarrow}}=0.5$ for panel (b).}
	\label{fig:ClassicalMotion}
\end{figure*}

\begin{figure}[th]
	\centerline{\includegraphics[width=0.5\textwidth]{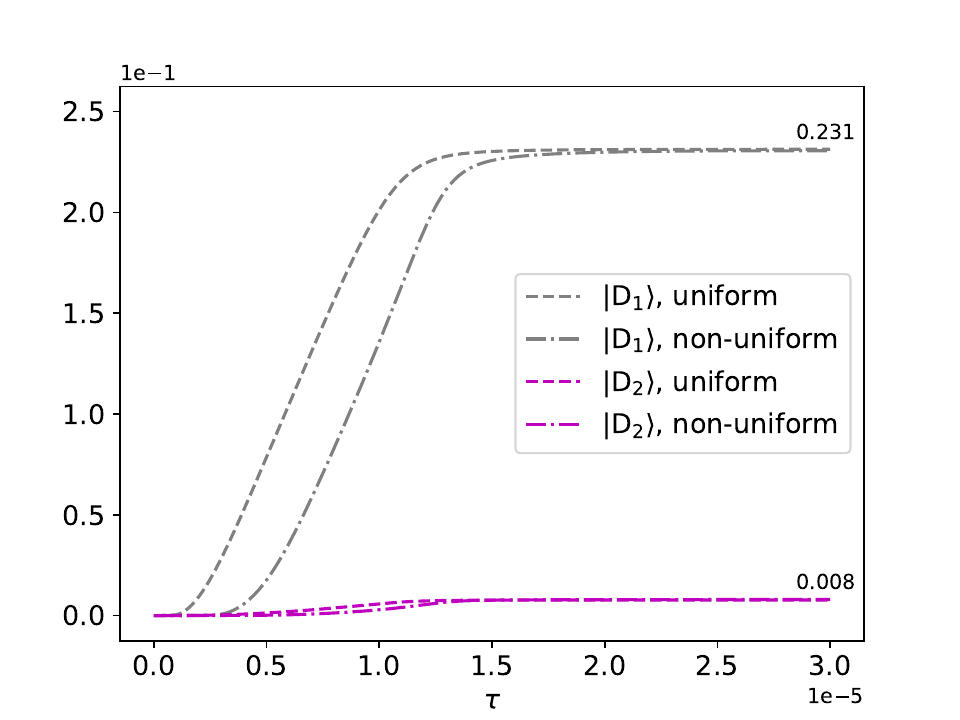}}
	\vspace*{8pt}
	\caption{(Color online) Dark states in the classical motion regime during dissociation. The same singlet dark state combinations as in the quantum case are identified: $|\text{D}_1\rangle = \frac{1}{\sqrt{2}}(|\Psi_0\rangle-|\Psi_1\rangle)$ and $|\text{D}_2\rangle = \frac{1}{\sqrt{2}}(|\Psi_2\rangle-|\Psi_3\rangle)$. Their populations during dissociation are $0.231$ and $0.008$, respectively (compared to $0.214$ and $0.022$ in the quantum motion case, Fig. \ref{fig:DarkStatesQuanMotion}). This indicates that dark state populations are slightly affected by the nuclear motion regime.}
	\label{fig:DarkStatesClassMotion}
\end{figure}

In Fig. \ref{fig:ClassicalMotion}, classical mechanical motion of the nuclei is considered. Except for the interaction forces, all other parameters are the same as those in Sec. \ref{subsec:ResultsQuantMotion}.
	
In panel (a) of Fig. \ref{fig:ClassicalMotion}, the initial state is also $|\Psi_{\text{initial}}\rangle$, representing two atoms in different cavities. For the association process, we consider two types of nuclear motion: uniform motion (Eq. \eqref{eq:StraightAss}) and non-uniform motion (Eq. \eqref{eq:TrigonometricAss}). The parameters are set to $\mu_{\Omega^{\uparrow,\downarrow,s}}=0.5$ and $\mu_{\omega^{\uparrow,\downarrow}}=0$, which promotes the formation of H$_2$. The numerical results in Fig. \ref{fig:ClassicalMotion} (a) show that both the dashed (uniform) and dash-dotted (non-uniform) red curves, representing the $|\mathrm{H}_2\rangle$ population, rise and approach unity at $\tau = T$. Although the two curves do not overlap, they converge to the same final value. However, the dashed curve rises faster than the dash-dotted curve, indicating that uniform motion leads to faster molecular formation. This difference arises from the different time dependence of the coupling strengths: linear change in the uniform case vs. trigonometric change in the non-uniform case.
	
In panel (b) of Fig. \ref{fig:ClassicalMotion}, the initial state is also $|\Psi_{\text{initial}}'\rangle$, representing a hydrogen molecule in a single cavity. For the dissociation process, we consider two types of nuclear motion: uniform motion (Eq. \eqref{eq:StraightDiss}) and non-uniform motion (Eq. \eqref{eq:TrigonometricDiss}), with parameters $\mu_{\Omega^{\uparrow,\downarrow,s}}=0$ and $\mu_{\omega^{\uparrow,\downarrow}}=0.5$ to promote dissociation. The results in Fig. \ref{fig:ClassicalMotion} (b) show that both the blue curves ($|\mathrm{H},\mathrm{H}\rangle$) and green curves ($|\mathrm{H}^-,\mathrm{H}^+\rangle$) converge to their asymptotic values ($\approx 0.51$ and $\approx 0.49$, respectively) at $\tau = T$. As in panel (a), uniform motion (dashed curves) yields a faster initial rise than non-uniform motion (dash-dotted curves). The dark state populations under classical motion are shown in Fig. \ref{fig:DarkStatesClassMotion}.

\section{Conclusion and future work} 
\label{sec:ConcluFuture}
	
In this paper, we have systematically compared quantum and classical nuclear motion in a finite-dimensional QED model of $\text{H}_2$ association and dissociation. 

The main novelty of this work is twofold. First, this work provides the first quantitative comparison between quantum tunneling and classical nuclear motion (described by time-dependent coupling strengths) in a modified Tavis--Cummings--Hubbard model for hydrogen chemistry. Second, two singlet dark states $|\text{D}_1\rangle = \frac{1}{\sqrt{2}}(|\Psi_0\rangle-|\Psi_1\rangle)$ and $|\text{D}_2\rangle = \frac{1}{\sqrt{2}}(|\Psi_2\rangle-|\Psi_3\rangle)$ are identified, which are decoupled from specific dissipation channels.

The key conclusions of this work are as follows:
\begin{itemize}
    	\item Time-scale separation: Classical motion reaches the same final states ($|\mathrm{H}_2\rangle$, $|\mathrm{H},\mathrm{H}\rangle$, $|\mathrm{H}^-,\mathrm{H}^+\rangle$) as quantum tunneling but requires an order-of-magnitude shorter time (Figs. 3 and 7).
    	\item Distinct population dynamics: In the quantum tunneling regime, final-state populations rise with an initially large slope that gradually decreases; in the classical regime, the slope is initially small, then large, then small again (Figs. 3 and 7).
    	\item Dark state populations: The dark state populations are $0.107$ and $0.011$ for quantum motion (Fig. 4) vs. $0.116$ and $0.004$ for classical motion (Fig. 8), indicating a subtle dependence on the nuclear motion regime.
    	\item Branching ratio: The dark states modify the branching ratio between $|\mathrm{H},\mathrm{H}\rangle$ and $|\mathrm{H}^-,\mathrm{H}^+\rangle$, explaining why $|\mathrm{H},\mathrm{H}\rangle$ has a higher population ($\approx 0.54$ for quantum, $\approx 0.51$ for classical) than $|\mathrm{H}^-,\mathrm{H}^+\rangle$ ($\approx 0.46$ for quantum, $\approx 0.49$ for classical).
\end{itemize}

Despite its current limitations, our approach offers the distinct benefits of simplicity and scalability. These findings contribute to a more complete understanding of the neutral hydrogen molecule association--dissociation model, thereby establishing a foundation for investigating more complex chemical and biological systems. Future work will extend this model to include non-Markovian environments and polyatomic systems.

\section*{Acknowledgments}
This work was supported by the China Scholarship Council (CSC No.202108090483).

\appendix

\section{Rotating wave approximation}
\label{appx:RWA}
	
The RWA is taken into account in this paper. When the strength of the applied electromagnetic radiation is close to resonance with an atomic transition and the strength is low, this approximation holds true \cite{Wu2007}. Thus,
\begin{equation}
	\label{appxeq:RWACondition}
	\frac{g}{\hbar\omega_{\text{cavity}}}\approx\frac{g}{\hbar\omega_{\text{atom}}}\ll 1,
\end{equation}
where $\omega_{\text{cavity}}$ stands for the cavity frequency and $\omega_{\text{atom}}$ for the transition frequency (of atom). The RWA allows us to change $\left(\sigma^{\dag}+\sigma\right)\left(a^{\dag}+a\right)$ to $\sigma^{\dag}a+\sigma a^{\dag}$. In Sec. \ref{subsec:Target} we assume that $\Omega_{\text{cavity}}=\Omega_{\text{atom}}$.

\section{Operators}
\label{appx:Operators}
	
For a $p$-photon state, the photon annihilation and creation operators are described as follows:
\begin{equation}
	\label{appxeq:PhotonOperators}
	\begin{aligned}
		&\text{if}\ p>0,\ \left\{
			\begin{aligned}
			&a|p\rangle=\sqrt{p}|p-1\rangle,\\
			&a^{\dag}|p\rangle=\sqrt{p+1}|p+1\rangle,
			\end{aligned}
			\right.\\
			&\text{if}\ p=0,\ \left \{
			\begin{aligned}
				&a|0\rangle=0,\\
				&a^{\dag}|0\rangle=|1\rangle.
			\end{aligned}
			\right.
	\end{aligned}
\end{equation}
The operators $a_{\Omega^{\uparrow,\downarrow,s}}$ and $a_{\omega^{\uparrow,\downarrow}}$, along with their Hermitian conjugates, all obey this rule. The interaction of the molecule with the electromagnetic field of the cavity, emitting or absorbing a photon of mode $\omega^{\uparrow,\downarrow}$, is described as follows:
\begin{equation}
	\label{appxeq:InteractionMolecule}
	\begin{aligned}
		&\sigma_{\omega^{\uparrow,\downarrow}}|1\rangle_{\Phi_1}^{\uparrow,\downarrow}|0\rangle_{\Phi_0}^{\uparrow,\downarrow}=|0\rangle_{\Phi_1}^{\uparrow,\downarrow}|1\rangle_{\Phi_0}^{\uparrow,\downarrow},\\
		&\sigma_{\omega^{\uparrow,\downarrow}}^{\dag}|0\rangle_{\Phi_1}^{\uparrow,\downarrow}|1\rangle_{\Phi_0}^{\uparrow,\downarrow}=|1\rangle_{\Phi_1}^{\uparrow,\downarrow}|0\rangle_{\Phi_0}^{\uparrow,\downarrow}.
	\end{aligned}
\end{equation}
The interaction of the atom with the electromagnetic field, emitting or absorbing a photon of mode $\Omega^{\uparrow,\downarrow}$, is described as follows:
\begin{equation}
	\label{appxeq:InteractionAtom}
	\begin{aligned}
		&\sigma_{\Omega^{\uparrow,\downarrow},i}|1\rangle_{\substack{\mathrm{at}_i\\\mathrm{or}_0}}^{\uparrow,\downarrow}|0\rangle_{\substack{\mathrm{at}_i\\\mathrm{or}_{-1}}}^{\uparrow,\downarrow}=|0\rangle_{\substack{\mathrm{at}_i\\\mathrm{or}_0}}^{\uparrow,\downarrow}|1\rangle_{\substack{\mathrm{at}_i\\\mathrm{or}_{-1}}}^{\uparrow,\downarrow},\\
		&\sigma_{\Omega^{\uparrow,\downarrow},i}^{\dag}|0\rangle_{\substack{\mathrm{at}_i\\\mathrm{or}_0}}^{\uparrow,\downarrow}|1\rangle_{\substack{\mathrm{at}_i\\\mathrm{or}_{-1}}}^{\uparrow,\downarrow}=|1\rangle_{\substack{\mathrm{at}_i\\\mathrm{or}_0}}^{\uparrow,\downarrow}|0\rangle_{\substack{\mathrm{at}_i\\\mathrm{or}_{-1}}}^{\uparrow,\downarrow}.
	\end{aligned}
\end{equation}
The interaction of the atom with the electromagnetic field, emitting or absorbing a photon of mode $\Omega^s$, is described as follows:
\begin{equation}
	\label{eq:InteractionAtomSpin}
	\begin{aligned}
		&\sigma_{\Omega^s,i}|1\rangle_{\substack{\mathrm{at}_i\\\mathrm{or}_0}}^{\uparrow}|0\rangle_{\substack{\mathrm{at}_i\\\mathrm{or}_0}}^{\downarrow}=|0\rangle_{\substack{\mathrm{at}_i\\\mathrm{or}_0}}^{\uparrow}|1\rangle_{\substack{\mathrm{at}_i\\\mathrm{or}_0}}^{\downarrow},\\
		&\sigma_{\Omega^s,i}|1\rangle_{\substack{\mathrm{at}_i\\\mathrm{or}_{-1}}}^{\uparrow}|0\rangle_{\substack{\mathrm{at}_i\\\mathrm{or}_{-1}}}^{\downarrow}=|0\rangle_{\substack{\mathrm{at}_i\\\mathrm{or}_{-1}}}^{\uparrow}|1\rangle_{\substack{\mathrm{at}_i\\\mathrm{or}_{-1}}}^{\downarrow},\\
		&\sigma_{\Omega^s,i}^{\dag}|0\rangle_{\substack{\mathrm{at}_i\\\mathrm{or}_0}}^{\uparrow}|1\rangle_{\substack{\mathrm{at}_i\\\mathrm{or}_0}}^{\downarrow}=|1\rangle_{\substack{\mathrm{at}_i\\\mathrm{or}_0}}^{\uparrow}|0\rangle_{\substack{\mathrm{at}_i\\\mathrm{or}_0}}^{\downarrow},\\
		&\sigma_{\Omega^s,i}^{\dag}|0\rangle_{\substack{\mathrm{at}_i\\\mathrm{or}_{-1}}}^{\uparrow}|1\rangle_{\substack{\mathrm{at}_i\\\mathrm{or}_{-1}}}^{\downarrow}=|1\rangle_{\substack{\mathrm{at}_i\\\mathrm{or}_{-1}}}^{\uparrow}|0\rangle_{\substack{\mathrm{at}_i\\\mathrm{or}_{-1}}}^{\downarrow}.
	\end{aligned}
\end{equation}
The tunneling operators of the nuclei, $\sigma_n$ and $\sigma_n^{\dag}$, have the following form:
\begin{equation}
	\label{appxeq:TunnelingOperators}
	\begin{aligned}
		&\sigma_n|1\rangle_n=|0\rangle_n,\\
		&\sigma_n^{\dag}|0\rangle_n=|1\rangle_n.
	\end{aligned}
\end{equation}

\section{Precise time step integration method}
\label{appx:PTSIM}
	
For decades, a dozen different methods for computing the matrix exponential have been examined \cite{Moler1978, Moler2003}. However, the fact that the exponential problem has not yet been fully resolved illustrates its importance. The PTSIM of matrix exponential was proposed in 1991 \cite{Zhong1991} and has been developed over the decades since then \cite{ZhongWilliams1994, Zhong1994, Zhong1995, Gao2016}. This method avoids computer truncation errors caused by fine division and improves the numerical solution of the matrix exponential to computer accuracy. When solving the equations of motion using stepwise integration, the matrix exponential operation involved is $e^{A\Delta t}$. According to the addition theorem of matrix exponentials:
\begin{equation}
	\label{eq:MatrixExp}
	\begin{aligned}
		e^{A\Delta t}&=\left[e^{A\frac{\Delta t}{2^N}}\right]^{2^N}\\
		&=\left[e^{A\varepsilon}\right]^{2^N},
	\end{aligned}
\end{equation}
where $A$ is a matrix, $\Delta t$ is the time step, and $\varepsilon=\frac{\Delta t}{2^N}$ (usually $N$ is taken to be equal to 20). The matrix exponential can be approximated using a Taylor series expansion (retaining four terms) as follows:
\begin{equation}
	\label{eq:Taylor}
	e^{A\varepsilon}\approx I+A\varepsilon+\frac{(A\varepsilon)^2}{2!}+\frac{(A\varepsilon)^3}{3!}+\frac{(A\varepsilon)^4}{4!},
\end{equation}
where $I$ is the unit matrix. Then,
\begin{equation}
	\label{eq:PTSIM}
	\begin{aligned}
		e^{A\Delta t}&\approx \left[I+A\varepsilon+\frac{(A\varepsilon)^2}{2!}+\frac{(A\varepsilon)^3}{3!}+\frac{(A\varepsilon)^4}{4!}\right]^{2^N}\\
		&=\left[I+T_{a,0}\right]^{2^N},
	\end{aligned}
\end{equation}
where $T_{a,0}=A\varepsilon+\frac{(A\varepsilon)^2}{2!}+\frac{(A\varepsilon)^3}{3!}+\frac{(A\varepsilon)^4}{4!}$. Then,
\begin{equation}
	\label{eq:Recursion}
	\begin{aligned}
		\left[I+T_{a,0}\right]^{2^N}&=\left[I+2T_{a,0}+T_{a,0}T_{a,0}\right]^{2^{N-1}}\\
		&=\left[I+T_{a,1}\right]^{2^{N-1}}\\
		&=\left[I+T_{a,2}\right]^{2^{N-2}}\\
		&=\cdots\\
		&=\left[I+T_{a,N}\right],
	\end{aligned}
\end{equation}
where $T_{a,n}=2T_{a,n-1}+T_{a,n-1}T_{a,n-1},\ n\geq 1$.

\bibliography{bibliography}
	
\end{document}